\documentclass[aip,jcp,twocolumn,superscriptaddress]{revtex4}%
\usepackage{epsfig,dsfont,amssymb,amsmath,amsthm,amsfonts,amsbsy,mathrsfs}
\usepackage{graphicx}
\usepackage{amsmath}
\usepackage{amssymb}%
\begin{document}
\title{Electronic Structures of Hybrid Graphene/Phosphorene Nanocomposite}

\author{Wei Hu}
\thanks{Corresponding author. E-mail: whu@lbl.gov}
\affiliation{Computational Research Division, Lawrence Berkeley
National Laboratory, Berkeley, CA 94720, USA} \affiliation{Hefei
National Laboratory for Physical Sciences at Microscale, University
of Science and Technology of China, Hefei, Anhui 230026, China}

\author{Tian Wang}
\affiliation{Department of Precision Machinery and Precision
Instrumentation, University of Science and Technology of China,
Hefei, Anhui 230026, China}


\author{Jinlong Yang}
\thanks{Corresponding author. E-mail: jlyang@ustc.edu.cn}
\affiliation{Hefei National Laboratory for Physical Sciences at
Microscale, University of Science and Technology of China, Hefei,
Anhui 230026, China}
\affiliation{Synergetic Innovation Center of
Quantum Information and Quantum Physics, University of Science and
Technology of China, Hefei, Anhui 230026, China}

\date{\today}

\pacs{ }

\begin{abstract}

Combining the electronic structures of two-dimensional monolayers in
ultrathin hybrid nanocomposites is expected to display new
properties beyond their simplex components. Here, first-principles
calculations are performed to study the structural, electronic and
optical properties of hybrid graphene and phosphorene nanocomposite.
It turns out that weak van der Waals interactions dominate between
graphene and phosphorene with their intrinsic electronic properties
preserved. Hybrid graphene and phosphorene nanocomposite shows
tunable band gaps at graphene's Dirac point and a transition from
hole doing to electron doing for graphene as the interfacial
distance decreases. Charge transfer between graphene to phosphorene
induces interfacial electron-hole pairs in hybrid graphene and
phosphorene nanocomposite with enhanced visible light response.

\end{abstract}

\maketitle

\section{Introduction}

Two-dimensional (2D) ultrathin materials, such as
graphene,\cite{Scinece_306_666_2004, NatureMater_6_183_2007}
silicene,\cite{PRL_102_236804_2009, SSR_67_1_2012} hexagonal boron
nitride,\cite{NatureMater_3_404_2004, NanoLett_10_3209_2010}
graphitic carbon nitride,\cite{NatureMater_8_76_2009,
JPCL_3_3330_2012} graphitic zinc oxide,\cite{JMC_15_139_2005,
PRL_96_066102_2006} molybdenum disulphide,\cite{PRL_105_136805_2010,
NatureNanotechnol_6_147_2011} have received considerable interest
recently owing to their outstanding properties and potential
applications. Graphene,\cite{NatureMater_6_183_2007} a 2D
sp$^2$-hybridized carbon monolayer, is known to have remarkable
electronic properties, such as a high carrier mobility, but the
absence of a bandgap limits its applications of large-off current
and high on-off ratio for graphene-based electronic devices.
Furthermore, intrinsic electronic properties of graphene depend
sensitively on the substrates due to strong graphene-substrate
interactions, such as SiO$_2$,\cite{NatPhys_4_144_2008,
PRL_106_106801_2011} SiC\cite{NatureMater_6_770_2007,
PRL_108_246104_2012} and metal\cite{PRL_101_026803_2008,
NatureNanotech_6_179_2011} surfaces. Therefore, opening a small
bandgap and finding an ideal substrate for graphene remains
challenging in the experiments.

Interestingly, many 2D ultrathin hybrid graphene-based
nanocomposites have been widely studied experimentally and
theoretically, such as graphene/silicene
(G/S),\cite{JCP_139_154704_2013, PRB_88_245408_2013,
APL_103_261904_2013} graphene/graphitic boron nitride
(G/g-BN),\cite{Nature_5_722_2010, NatureMater_10_282_2011,
PRB_81_155433_2010} graphene/graphitic carbon nitride
(G/g-C$_3$N$_4$),\cite{JPCC_115_7355_2011, JACS_133_8074_2011,
JACS_134_4393_2012}, graphene/graphitic zinc oxide
(G/g-ZnO)\cite{JCP_138_124706_2013, JPCC_117_10536_2013,
JAP_113_054307_2013} and graphene/molybdenum disulphide
(G/MoS$_2$)\cite{Nanoscale_3_3883_2011, JPCC_117_15347_2013,
Science_340_1311_2013} These hybrid graphene-based nanocomposites
show much more new properties far beyond their simplex components.
Furthermore, most of them are ideal substrates for graphene to
preserve the intrinsic electronic properties of graphene and
substrates.

Most recently, a new 2D material, namely, black phosphorus monolayer
or phosphorene,\cite{PRB_86_035105_2012, JPCC_118_14051_2014,
NatureNanotech_9_372_2014, NatureCommun_5_4475_2014} has been
isolated in the experiments through mechanical exfoliation from bulk
black phosphorus and has immediately received considerable
attention. Phosphorene also shows some remarkable electronic
properties superior to graphene. For example, phosphorene is a
semiconductor with a direct bandgap of about 1
$eV$,\cite{JPCC_118_14051_2014} showing the drain current modulation
up to 10$^5$ and carrier mobility up to 10$^3$ cm$^2$/(Vs) in
nanoelectronics.\cite{NatureNanotech_9_372_2014} Here, an
interesting question arise: whether graphene and phosphorene can
form a 2D hybrid G/P nanocomposite with new properties?

In the present work, we design a new 2D hybrid graphene and
phosphorene nanocomposite and study its electronic and optical
properties with first-principles calculations. The results show that
graphene interacts overall weakly with phosphorene via van der Waals
(vdW) interactions, thus, their intrinsic electronic properties can
be preserved in hybrid graphene/phosphorene nanocomposite. Moreover,
interlayer interactions in hybrid graphene/phosphorene nanocomposite
can induce tunable band gaps at graphene's Dirac point, a transition
from hole doing to electron doing for graphene and enhanced visible
light response.

\section{Theoretical Methods and Models}

The lattice parameters of graphene and phosphorene calculated to
setup unit cell are $a$(G) = $b$(G) = 2.47
{\AA},\cite{JCP_139_154704_2013} $a$(P) = 4.62 {\AA} and $b$(P) =
3.30 {\AA}.\cite{JPCC_118_14051_2014} We design a new 2D hybrid
graphene/phosphorene nanocomposite (40 carbon atoms and 28
phosphorus atoms) as shown in Figure 1 with a small lattice mismatch
less than 2\%. The vacuum space in the Z direction is about 15 {\AA}
to separate the interactions between neighboring slabs.

\begin{figure}[htbp]
\begin{center}
\includegraphics[width=0.5\textwidth]{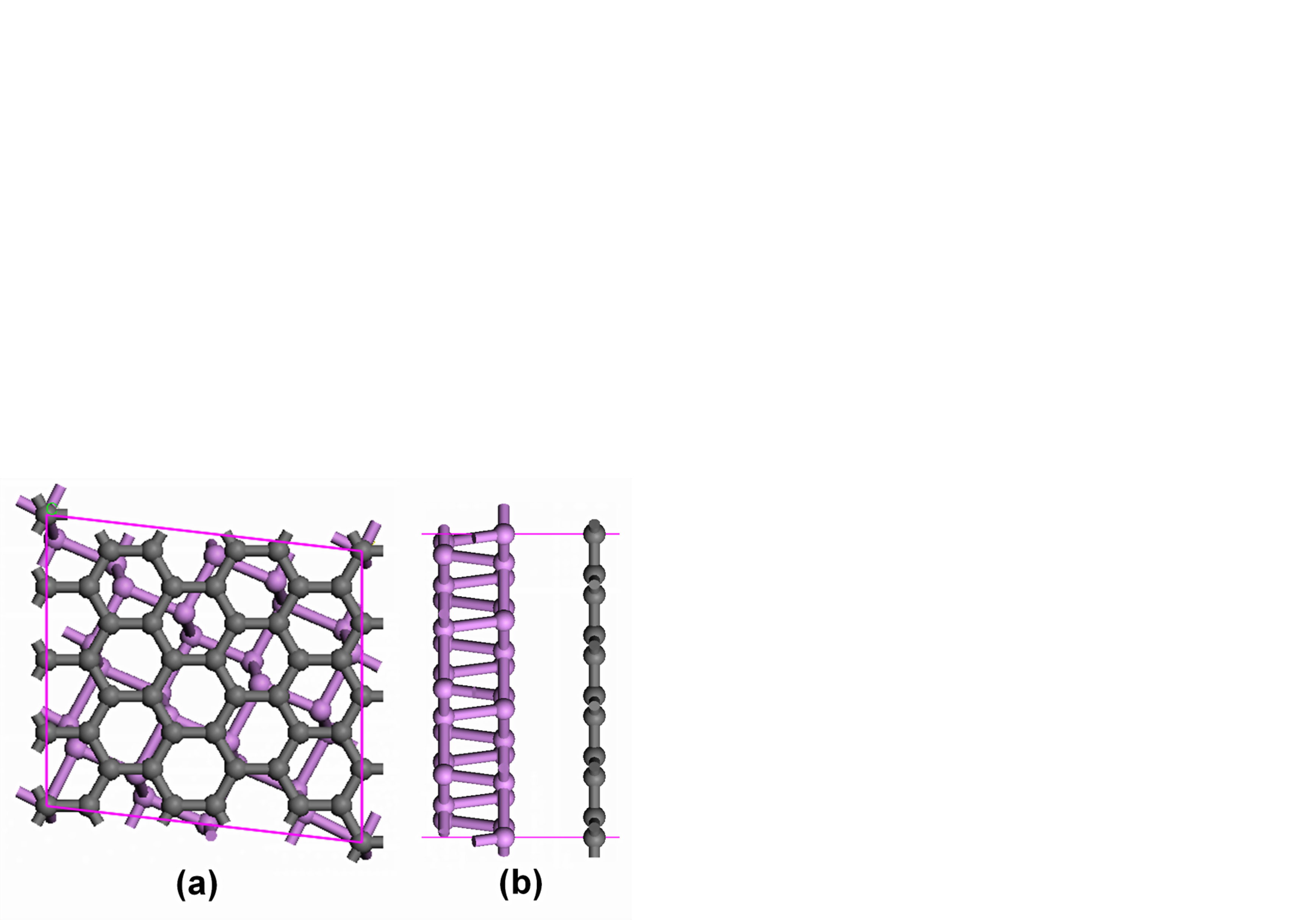}
\end{center}
\caption{(Color online) Geometric structures of hybrid
graphene/phosphorene nanocomposite ((a) top and (b) side views). The
gray and violet balls denote carbon and phosphorus atoms,
respectively.}
\end{figure}

First-principles calculations are based on the density functional
theory (DFT) implemented in the VASP
package.\cite{PRB_47_558_1993_VASP} The generalized gradient
approximation of Perdew, Burke, and Ernzerhof
(GGA-PBE)\cite{PRL_77_1996} with van der Waals (vdW) correction
proposed by Grimme (DFT-D2)\cite{JCC_27_1787_2006_Grimme} is chosen
due to its good description of long-range vdW
interactions.\cite{JPCC_111_11199_2007, PCCP_10_2722_2008,
NanoLett_11_5274_2011, PRB_83_245429_2011, PRB_85_125415_2012,
PRB_85_235448_2012, JPCL_4_2158_2013, PCCP_15_5753_2013,
Nanoscale_5_9062_2013} As an benchmark, DFT-D2 calculations give a
good bilayer distance of $c$ = 3.25 ${\AA}$ and binding energy of
E$_b$ = -25 $meV$ per carbon atom for bilayer graphene, which fully
agree with previous experimental\cite{PR_100_544_1955,
PRB_69_155406_2004} and theoretical\cite{PRB_85_205402_2012,
JCP_138_054701_2013} studies. The energy cutoff is set to be 500
$eV$. The surface Brillouin zone is sampled with a 3 $\times$ 3
regular mesh and 120 $k$ points are used for calculating the tiny
band gaps at the Dirac point graphene in the hybrid G/P
nanocomposite supercell. All the geometry structures are fully
relaxed until energy and forces are converged to 10$^{-5}$ $eV$ and
0.01 $eV$/{\AA}, respectively.

To investigate the optical properties of hybrid G/P nanocomposite,
the frequency-dependent dielectric matrix is
calculated.\cite{PRB_73_045112_2006} In order to calculate the
optical properties of hybrid G/P nanocomposite, a large 6 $\times$ 6
regular mesh for the surface Brillouin zone, a large number of empty
conduction band states (two times more than the number of valence
band) and frequency grid points (2000) are adopted. We crosscheck
the optical properties of graphene and phosphorene, consistent with
previous theoretical calculations.\cite{JCP_139_154704_2013,
NatureCommun_5_4475_2014}


\section{Results and Discussion}

Electronic properties of pristine graphene and phosphorene
monolayers in the supercells are checked first and their electronic
band structures are plotted in Figure 2a and 2b. Graphene is a
zero-gap semiconductor (Figure 2a), showing a linear Dirac-like
dispersion relation $E$($k$) = $\pm$$\hbar$$\nu_{F}$$|$$k$$|$ around
the Fermi level where $\nu$$_{F}$ is the Fermi velocity, and
$\nu$$_{F}$$(G)$ = 0.8$\times$10$^6$ $m/s$ at the Dirac point of
graphene. Monolayer phosphorene is semiconducting with a direct band
gap of 0.85 $eV$ (Figure 2b), which agrees well with previous
theoretical studies.\cite{JPCC_118_14051_2014}

\begin{figure}[htbp]
\begin{center}
\includegraphics[width=0.5\textwidth]{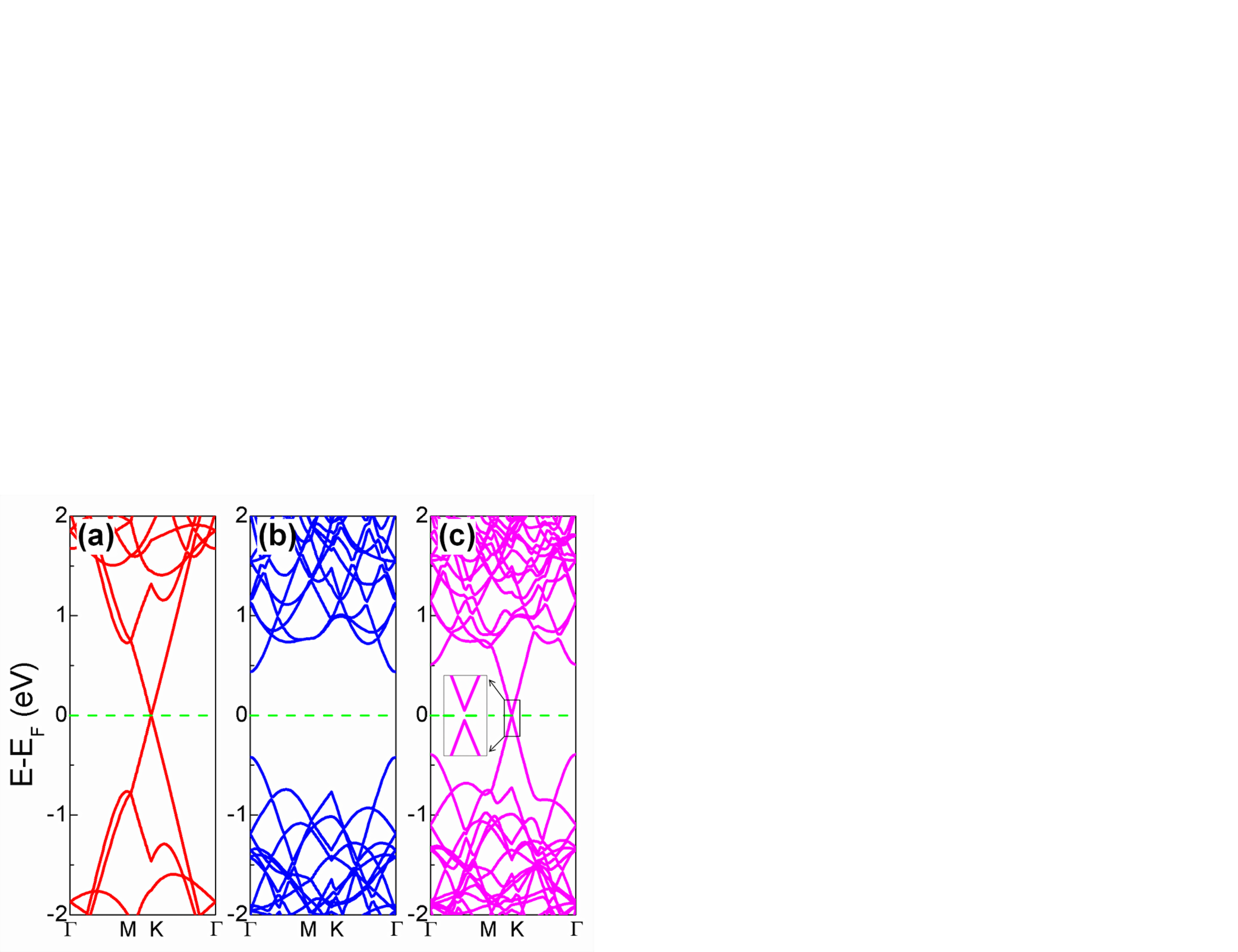}
\end{center}
\caption{(Color online) Electronic band structures of (a) graphene,
(b) phosphorene and (c) hybrid graphene/phosphorene nanocomposite.
The Fermi level is set to zero and marked by green dotted lines.}
\end{figure}

We then study the geometric structures of hybrid G/P nanocomposite.
Typical vdW equilibrium spacing of about 3.43 {\AA} with
corresponding small binding energy of about -24.7 $meV$ per atom of
graphene is obtained for hybrid G/P nanocomposite, which is well
comparable with recent theoretical calculations in 2D graphene based
nanocomposites, such as G/S,\cite{JCP_139_154704_2013}
G/g-BN,\cite{PRB_81_155433_2010}
G/g-C$_3$N$_4$,\cite{JACS_134_4393_2012}
G/g-ZnO\cite{JCP_138_124706_2013} and
G/MoS$_2$.\cite{Nanoscale_3_3883_2011} Thus, weak vdW interactions
dominate between graphene and phosphorene, suggesting that
phosphorene can be used as an ideal substrate for graphene with
their intrinsic electronic structures undisturbed. Notice that the
small lattice mismatch of about 2\% for graphene and phosphorene has
little effect on their electronic properties in hybrid G/P
nanocomposite.

Electronic band structure of hybrid G/P nanocomposite is shown in
Figure 2c. The Dirac point of graphene is still preserved, and the
Fermi velocity at the Dirac point is almost unchanged
($\nu$$_{F}$$(G/P)$ = 0.8$\times$10$^6$ $m/s$) in hybrid G/P
nanocomposite compared to free-standing graphene, though small band
gap of 10 $meV$ is opened at the Dirac point of grpahene. Notice
that induced band gaps at the Dirac point of graphene are typically
sensitive and tunable to other external conditions, such as
interlayer separation,\cite{JCP_139_154704_2013} as plotted in
Figure 3. The gap values at graphene's Dirac point increase from 5
to 90 $meV$ as the interfacial distance decreases from 3.7 to 2.7
{\AA} in hybrid G/P nanocomposite, showing a potential for
high-performance graphene-based electronic devices.

\begin{figure}[htbp]
\begin{center}
\includegraphics[width=0.5\textwidth]{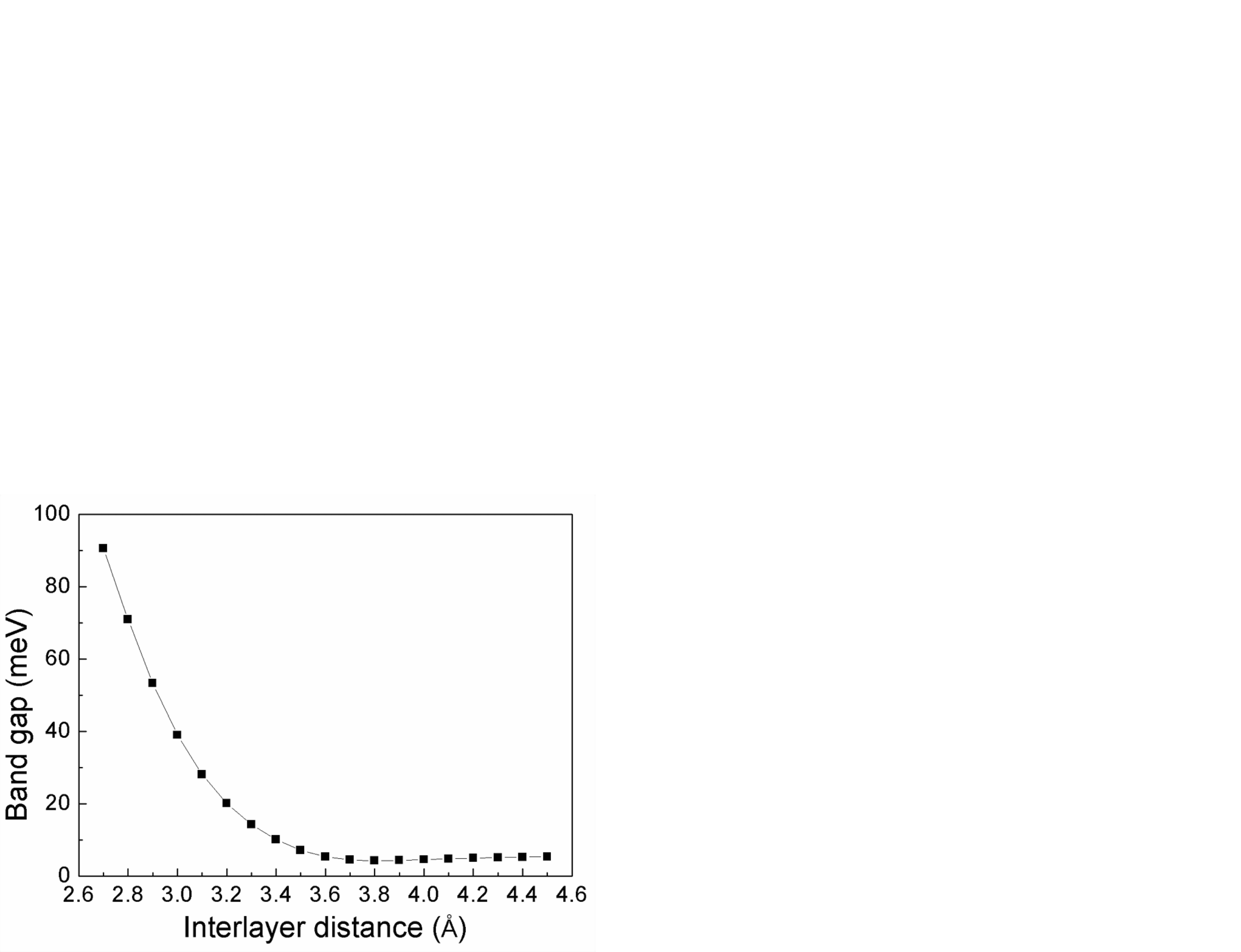}
\end{center}
\caption{(Color online) The band gap opened at graphene's Dirac
point in hybrid graphene/phosphorene nanocomposite as a function of
interfacial distance.}
\end{figure}

Interestingly, we find that the Dirac point of graphene moves form
the conduction band to the valence band of phosphorene as the
interfacial distance decreases from 4.5 to 2.7 {\AA} in hybrid G/P
nanocomposite, inducing a transition from hole (p-type) doping to
electron (n-type) doping for graphene as shown in Figure 4. When the
interfacial distance artificially increases larger than 4.5 {\AA},
graphene's Dirac point is close to phosphorene's conduction band.
That is because graphene's work function (4.3 $eV$) is close to
phosphorene's nucleophilic potential (4.1 $eV$). Based on the
Schottky-Mott model,\cite{PR_71_717_1947} electrons easily transfer
from graphene to phosphorene, resulting in p-type doping of
graphene. But, weak overlap of electronic states between graphene
and phosphorene are enhanced as the interfacial distance decreases.
Furthermore, carbon has a large electronegativity (2.55) than that
(2.19) of phosphorus. Thus, when graphene and phosphorene are close
to each other (smaller than 3.0 {\AA}), electrons easily transferred
from phosphorene to graphene, resulting in n-type doping of
graphene. Similarly, interlayer-interaction induced a transition
from p-type doping to n-type doping for graphene has also observed
experimentally and theoretically in hybrid graphene/silicene
nanocomposite\cite{JCP_139_154704_2013} and graphene adsorption on
some metal substrates.\cite{PRL_101_026803_2008}

\begin{figure}[htbp]
\begin{center}
\includegraphics[width=0.5\textwidth]{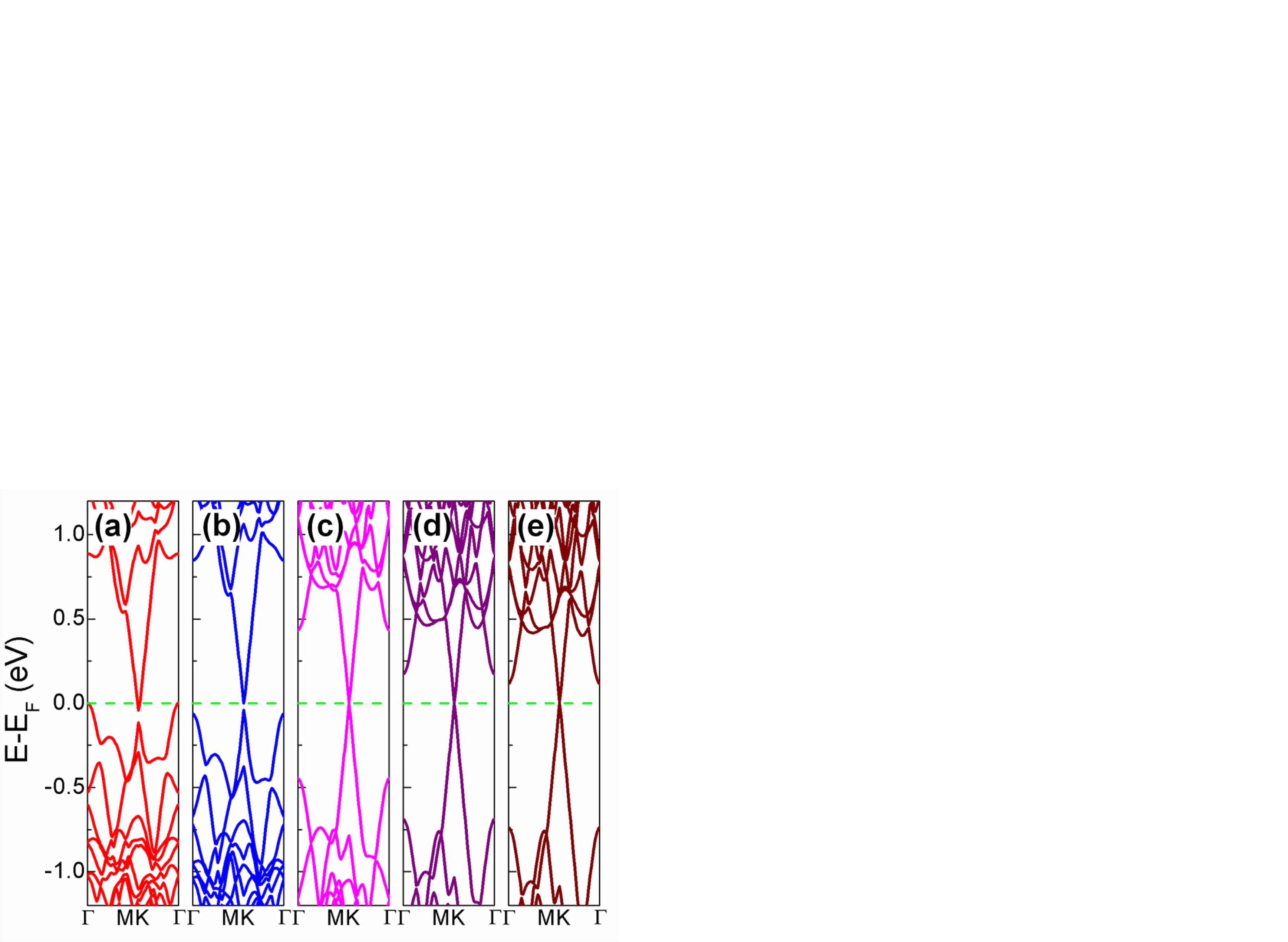}
\end{center}
\caption{(Color online) Electronic band structures of hybrid
graphene/phosphorene nanocomposite at different interfacial
distances $D$ = 2.8, 3.0, 3.5, 4.0 and 4.5 {\AA}. The Fermi level is
set to zero and marked by green dotted lines.}
\end{figure}

Besides commonly focused electronic structures in 2D graphene based
nanocomposites, we also study the optical properties in hybrid G/P
nanocomposite. Though pristine graphene and phosphorene themselves
display unique optical properties,\cite{JCP_139_154704_2013,
NatureCommun_5_4475_2014} interlayer interactions and charge
transfer in 2D nanocomposites may induce new optical
transitions.\cite{JPCL_3_3330_2012, JCP_138_124706_2013} As shown in
Figure 5, charge transfer between graphene to phosphorene induces
interfacial electron-hole pairs in hybrid G/P nanocomposite. In
optical property calculations, the imaginary part of dielectric
function for graphene and phosphorene monolayers as well as
corresponding hybrid G/P nanocomposite are evaluated, including the
light polarized parallel and perpendicular to the plane, as shown in
Figure 5. Parallel optical absorption of pristine graphene and
phosphorene monolayers mainly possesses in the visible light range
from 200 to 1000 $nm$ due to the transitions from $\pi$ to $\pi$$^*$
states and $\sigma$ to $\sigma$$^*$ states. Hybrid G/P nanocomposite
exhibits stronger optical absorption, especially in the visible
light range from 300 to 800 $nm$, compared with simplex graphene and
phosphorene monolayers, because the interlayer coupling in hybrid
G/P nanocomposite induces electronic states overlap and electrons
can now be directly excited between graphene and phosphorene. But,
perpendicular dielectric function is almost unaffected by the
interlayer interactions in hybrid G/P nanocomposite due to very weak
optical absorption of graphene in this direction.

\begin{figure}[htbp]
\begin{center}
\includegraphics[width=0.5\textwidth]{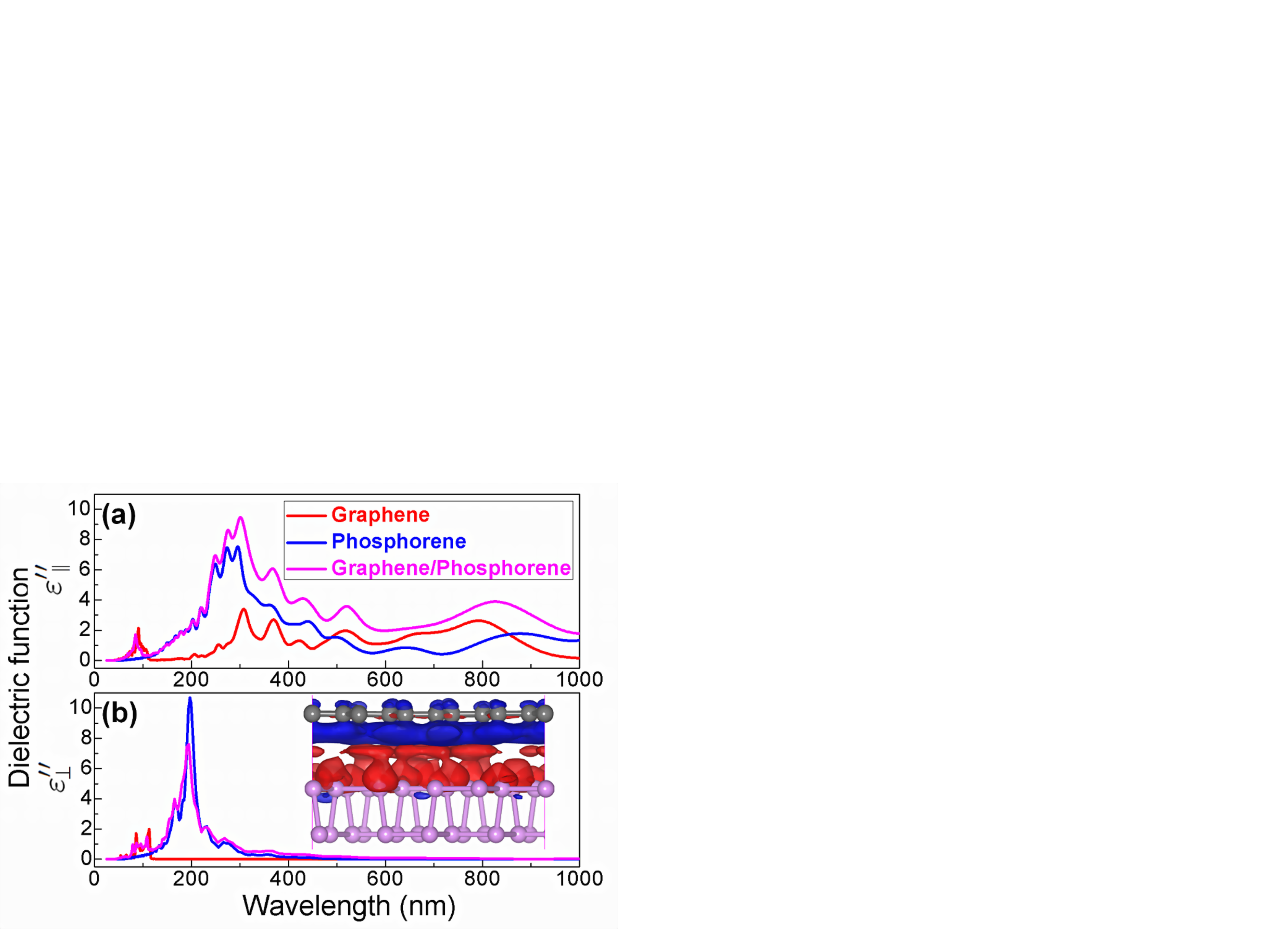}
\end{center}
\caption{(Color online) Imaginary part of frequency dependent
dielectric function ((a) parallel and (b) perpendicular) for
pristine graphene and phosphorene monolayers as well as
corresponding hybrid graphene/phosphorene nanocomposite.
Differential charge density (0.002 $e$/{\AA}$^3$) of hybrid
graphene/phosphorene nanocomposite is shown in the insert.}
\end{figure}

\section{Conclusions}

In summary, we study the electronic structures and optical
properties of hybrid graphene/phosphorene nanocomposite with
first-principles calculations. We find that phosphorene interacts
weakly with graphene via weak vdW interactions to preserve their
intrinsic electronic properties. Moreover, interlayer interactions
can induce tunable band gaps at graphene's Dirac point, a transition
from hole doing to electron doing for graphene and enhanced visible
light response in hybrid graphene/phosphorene nanocomposite. With
the excellent electronic and optical properties combined beyond
simplex graphene and phosphorene monolayers, 2D ultrathin hybrid
graphene/phosphorene nanocomposite system is expected to be of a
great potential in new graphene-based electronic devices.

\section{ACKNOWLEDGMENTS}

This work is partially supported by the National Key Basic Research
Program (2011CB921404), by NSFC (11404109, 21121003, 91021004,
21233007, 21222304), by CAS (XDB01020300). This work is also
partially supported by the Scientific Discovery through Advanced
Computing (SciDAC) program funded by U.S. Department of Energy,
Office of Science, Advanced Scientific Computing Research and Basic
Energy Sciences (W. H.). We thank the National Energy Research
Scientific Computing (NERSC) center, USTCSCC, SC-CAS, Tianjin, and
Shanghai Supercomputer Centers for the computational resources.

\end{document}